**Free-ranging dogs do not distinguish between barks without context**

Prothama Manna[1], Anindita Bhadra[1,*]

[1]Behaviour and Ecology Lab, Department of Biological Sciences,

Indian Institute of Science Education and Research Kolkata,

Mohanpur Campus, Mohanpur 741246

West Bengal, India

**\*Address for Correspondence:**

Behaviour and Ecology Lab, Department of Biological Sciences,

Indian Institute of Science Education and Research Kolkata

Mohanpur Campus, Mohanpur,

PIN 741246, West Bengal, India

tel.+91-33 6136 0000 ext 1223

fax+91-33-25873020

e-mail: abhadra@iiserkol.ac.in

**Abstract**

Canids display a vast diversity of social organizations, from solitary-living to pairs to packs. Domestic dogs have descended from pack-living gray wolf-like ancestors. Unlike their group living ancestors, free-ranging dogs are facultatively social, preferring to forage solitarily. They are scavengers by nature, mostly dependent on human garbage and generosity for their sustenance. Free-ranging dogs are highly territorial, often defending their territories using vocalizations. Vocal communication plays a critical role between inter and intraspecies and group interaction and maintaining their social dynamics. Barking is the most common among the different types of vocalizations of dogs. Dogs have a broad hearing range and can respond to sounds over long distances. Domestic dogs have been shown to have the ability to distinguish between barking in different contexts. Since free-ranging dogs regularly engage in various kinds of interactions with each other, it is interesting to know whether they are capable of distinguishing between vocalizations of their own and other groups. In this study, a playback experiment was used to test if dogs can distinguish between barking of their own group member from a non-group member. Though dogs respond to barking from other groups in territorial exchanges, they did not respond differently to the self and other group barking in the playback experiments. This suggests a role of context in the interactions between dogs and opens up possibilities for future studies on the comparison of the responses of dogs in playback experiments with their natural behavior through long-term observations.

## Introduction

Living in a group has considerable impact on an individual's life. Communication, both within and between groups is imperative for group living to be sustained. Vocalizations are primarily used in social interactions by higher order organisms, and tend to attract the attention of humans due to their similarity to the most common means of human communication – speech. Humans are probably the only species to have evolved a multitude of complex linguistic systems for communication. However, other species like honeybees, dolphins, elephants etc. have been found to use complex communication that are akin to languages (Seyfarth & Cheney, 2010). Communication in animals, however, is not limited to vocalizations and can be acoustic, visual, olfactory and tactile. How individuals communicate with conspecifics and with other species and how these systems vary from human communication systems, are questions that continue to intrigue us, leading to a large body of research (Marler P, 1961). While communication is a pre-requisite for evolving social behaviour, social interactions tend to shape the personalities of individuals, influencing the manner in which they communicate. For example, social interactions contextualize vocalizations (Yin, 2002) and may guide an individual's usage of and response to vocalization, playing an essential role in the individual's ability to communicate effectively (Rendall, Seyfarth, Cheney, & Owren, 1999);(De La Torre & Snowdon, 2002). The intricacies of vocal communication can be best studied in group living species that use various kinds of vocalizations in different social contexts.

Canids are good model systems for studying vocalizations, as they show different levels of social organization and actively communicate using vocalizations, though olfactory signals also play an important role in canid communication(Cohen & Fox, 1976). For example, within wolf packs, howling plays a critical role to reassemble separated individuals, as well as to communicate information on individual identity, location, and other behavioral and environmental factors (Theberge & Falls, 1967). Communication in dogs (*Canis lupus familiaris*) involves both their conspecifics and humans. Dogs have a broad range of vocal repertoire (Yeon, 2007). Although their vocalizations are quite similar to their close relatives, the gray wolf (*Canis lupus lupus*), dogs vocalize in a wider variety of social contexts as compared to wolves (Pongrácz, Molnár, & Miklósi, 2010). The vocal behaviour of dogs underwent considerable changes during the domestication process, which is considered to be a result of the dog's adaptation to the human social environment (Feddersen-Petersen, 2000). Among the different vocalizations, the bark is undoubtedly the most typical of dogs. Barks are the short and repetitive signals, with a highly variable acoustic structure (dominant frequency range between 160 and 2630 Hz), which also differs between breeds and even between individuals. They are generally used in short-range interactions and several behavioral contexts, like greeting, warning/alerting, calling for attention, or during play (Feddersen-Petersen, 2000). Recent studies have reported that similar to barks, growls also convey meaningful information to dogs (Taylor, Reby, & McComb, 2009). These low-frequency broadband vocalizations are mainly produced during agonistic interactions as a warning or threatening signal or play interactions. Dog's acoustic communication includes whines, which are indicators of stressful arousal but also greeting and attention-seeking behaviour; howls, which maintain group cohesion; groans, signs of acute distress and acute pain, respectively; and grunts, which are generally considered as a pleasure-related signal (Yeon, 2007).

Behavioral variation between wild and domestic populations is considered to be a reflection of change in genetic variation caused by the domestication process. This might hold true for domestic dogs (Yin, 2002). Though dogs are primarily perceived as domesticated animals or pets, nearly 80% of the world's dog population comprises of free-ranging dogs (Daniels & Bekoff, 2015). Domestic dogs which are not under direct human supervision and whose activities and movements are not restricted by humans are termed as free-ranging dogs (Serpell, 2016). Free-ranging dogs are distributed in most countries of the global south and are known to live in groups (Verardi, Lucchini, & Randi, 2006). They occupy every possible human habitation in India, from forest fringes to metropolitan areas, from the Himalayas to the coasts (Sen Majumder, Paul, Sau, & Bhadra, 2016). Pet dogs are typically under the supervision of their owners, typically deprived of their ancestral social environments during development; such social changes have the potential to lead to changes in their vocal habits. Free-ranging dogs, on the other hand, tend to live in stable social groups (Majumder et al., 2014), and they interact with other groups in situations of conflict like territory maintenance and sometimes of affiliation, like mating. They also show interesting cooperation-conflict dynamics within their groups () through various affiliative and agonistic interactions. Vocalization plays a vital role in maintaining the social dynamics within and between groups.

Pet dogs are known to have the ability to discriminate between conspecific barks emitted in different domestic contexts recorded either from the same or different individuals (Molnár, Pongrácz, Faragó, Dóka, & Miklósi, 2009). Using spectrogram analysis, dog barks can be divided into different subtypes based on their context and dogs can be identified by their bark spectrograms, irrespective of the context of the bark (Yin & McCowan, 2004). Several studies have been carried out to understand communication in dogs and between dogs and humans, involving pets, but little is understood of this behavior in a natural environment. It is a common observation that free-ranging dogs not only exchange barks and howls among their group members and neighbouring groups, but also with groups that are out of the visual range of the dogs. Hence vocalizations definitely play a key role in inter and intra-group communication. However, it is not understood whether the free-ranging dogs can identify and respond accordingly to the barks of individuals, without the context of the territory, by merely hearing the sound.

In this study, we carried out a set of behavioral observations and experiments to understand if they respond differently to the recorded barks of their own and other groups, in a playback experiment carried out within their territories. We hypothesized that if dogs are able to distinguish between the barks and identify their own from the other, they would respond differently to the two sounds. We thus aimed to use the response of the dogs, both as individuals and as groups, to the playback tracks as a bioassay for addressing the more interesting question – can dogs distinguish between their own barks and those of others, in the absence of any context.

**Methods-**

   A. **Subjects and study area**

In this study, we tested 157 free-ranging dogs belonging to 40 groups with a minimum group size of three adults. Individuals that were sighted to show affiliative interactions (allogrooming, playing, walking together, sharing food, etc.), resting or moving together, within an approximate distance of 1 meter from each other, and/or defending territories together against other dogs, were considered as a group. Dog groups were located in four different urban areas – Bandel

(22.9342° N, 88.3714° E), Chinsurah (22.9012° N, 88.3899° E), Balindi (22.9740° N, 88.5382° E) and Kalyani (22.9751° N, 88.4345° E) of West Bengal, India **(See Supplementary Figure 1 for details)** and tracked for several days to ensure group identity, prior to the experiment. For all the dogs used in the study, we recorded the sex (by observing their genitalia) and the age class (pups, juveniles or adults, based on body size and genital structures) of the dog (Morris D, 1987). For each of the groups, the members of the group were identified individually using a combination of coat colour patterns, other physical features like ear notching, bending of ears, scars etc. and sex.

### B. Experimental Procedure

We used a playback to test whether the dogs can distinguish between their own group's barking sound from that of a different group. The experiment consisted of two phases – the recording and the playback phases, which were carried out on two different days with a minimum gap of 30 days between the two phases **(See Supplementary Figure 2 for details)**.

**(i) The Recording phase**

The experimenter (E) went to the territory of each group and ensured that at least three adult individuals of the group were present, before carrying out the experiment. E carried out behavioural observations on the group for one hour using All Occurrences Sessions (AOS) of 10 minutes each, followed by two minutes breaks (Altmann, Loy, & Wagner, 1973). She recorded all the vocalizations of the group that occurred during this period using a Sony IC audio recorder ICD-UX533F irrespective of the context of the vocalization. Signals were recorded at a sample rate of 44.1 kHz and a sample size of 16 bits. For each group, the barking track which had the least noise and was of the longest duration, was selected after completion of observations. The chosen tracks were subjected to removal of additional background noise using the Ocenaudio software. The tracks thus prepared were used for the playback experiments.

**(ii) Playback phase**

Three barking tracks were used for the playback experiment, among which one was the self group's barking, and the other two were from two randomly chosen "other" groups. This phase of the experiment was further divided into the following phases:

a) **Observation phase I:** E went to the territory of each group and ensured that at least three adult individuals belonging to the group were present in the area. She hid a Bluetooth speaker within the territory, either behind a bush or some human artefact already existing within the territory, approximately 3-meter away from the center of the group. She then carried out observations using AOS as before, for one hour. This was done to check for the baseline behaviour of the group (control) before the test or playback phase.

b) **Playback phase:** Following the observation phase I, the experimenter played one of the three tracks, chosen randomly, from her cell phone using the Bluetooth speaker. When the track stopped, the group was allowed 2 minutes to settle down before playing the next track. All three tracks were played in random order during the playback phase I, with 2 minutes settling time between the tracks. Each track was played only once to a group. E recorded the response of the dogs on a Sony HDR-PJ230 video camera during this phase.

c) **Observation phase II:** At the end of the playback phase, the group was observed for 30 minutes using AOS as before. This was done to ensure that the group settled down after the playback phase.

The playback phase was repeated twice more, with a 30-minute observation phase in between them. Thus, there were three playback phases in total, interspersed with two observation phases of 30 minutes each. The three tracks were randomized for the order of playback during each of the playback phases.

### C. Data analysis and statistics

All the videos were coded by E, and the data subjected to further analysis. The responses of each group were coded for latency, response duration, the type of response, and the number of individuals that gave a response. The analysis was carried out for the tracks separately, grouped as self and other. The alpha level was 0.05 throughout the analysis. All statistical analyses were performed in R 3.5.3 (R Core Team, 2019).

(i) **Response index:** Responses were categorized into three major behavioural types – change of state, movement, and vocalization. All the vocalization responses were further categorized into vocalization 1 and vocalization 2, according to the experimenter's perception. The less energetic and aggressive vocalizations were grouped under vocalization 1 category, and the vocalization responses which were most energetic and aggressive were classified under vocalization 2 category. While a change of state was considered to be the least costly in terms of energy expenditure, vocalizations like angry bark, barking back, growl, etc. were expected to be the most energy intensive. Other vocalizations like bark, howl, whoop, etc. were considered to be intermediate in their energy demands, followed by movements like trotting, running, etc. When a dog showed a combination of two or three responses, it was categorized under the highest of the response categories shown at that time. A response index was estimated using these responses. The response categories were given a score of one to four according to the experimenter's view of energy expense by dogs (**Table 1**).

| Type of response | Score |
|---|---|
| Change of state | 1 |
| Movement | 2 |
| Vocalization 1 | 3 |
| Vocalization 2 | 4 |

**Table 1. Response index incorporating the type of response and their corresponding scores**

The responses in the self and other conditions were compared using contingency chi-square tests. The latency, response duration, the proportion of individuals that responded were compared using the Wilcoxon paired-sample tests across all the trials. To check for the gender bias of the responders in self and other conditions, a contingency chi-square test was performed. The percentage of the four kinds of responses shown in the self and other conditions were compared using a contingency chi-square test. The goodness of fit test was used to check for gender bias of the responding individuals and type of response on hearing any of the barks.

(ii) **Consistency of response:** In the playback phase, whenever there were at least two responses in the three trials for each condition (self and other), the observed consistency of the response of each group, individual and the first responder were compared with the expected level using contingency chi-square test. If a response was obtained in all the three trials, then it was categorized as high consistency, whereas two responses in any of the two trials were categorized as low consistency. A Goodness of fit test was used to check for gender bias of the first responder.

**Sound analysis:** Ten randomly chosen syllables from each track were analyzed for six different and most commonly reported acoustic parameters **(Table 2),** using Raven Pro 1.5 software (Bioacoustics Research Program, 2014) **(Figure 3).** Principal components analysis (PCA) was carried out using all the ten syllables of every group. This was done to check if the tracks were comparable in their acoustic signatures, irrespective of whether the dogs could identify them as "self" and "other." Spectrograms were made with a 512-point (11.6 ms) Hann window (3 dB bandwidth = 124 Hz), with 75% overlap, and a 1024-point DFT, yielding time and frequency measurement precision of 2.9 ms and 43 Hz.

| Parameters | Descriptions |
|---|---|
| High_f | The upper frequency bound of the selection. (Hz) |
| Low_f | The lower frequency bound of the selection. (Hz) |
| Peak_f | The frequency at which Max Power occurs within the syllable. (Hz) |
| Duration 90% | The difference between the 5% and 95% times. (s) |
| Bandwidth 90% | The difference between the 5% and 95% frequencies. (Hz) |
| Aggregate Entropy | The disorder in a sound by analyzing the energy distribution within a syllable. |

**Table 2. The acoustic parameters used in the spectrogram analysis and their description**

A generalized linear mixed model (GLMM) analysis was carried out to check for the effect of these parameters on the response index for the groups, Laplace approximation using the "glmer" function in the "lme4" package with group_ID as a random effect and response index as the fixed effect was used for the GLMM analysis. AIC values were compared in order to get the best-fitting models.

Further, Principal Components Analysis (PCA) was carried out using the ten syllables of all the groups, considering locality as a variable, and A generalized linear model (GLM) analysis was carried out with Poisson Regression to check for the effect of the locality and group size on the response index for the groups.

**(iii) Analysis of AOS data:** The frequency per hour of aggression, affiliation, urine marking and vocalization behaviours in the three AOS for each group in the playback experiment and the recording experiment were compared using Kruskal-Wallis rank sum tests followed by pairwise Mann-Whitney tests with Bonferroni correction method whenever required.

**Results**

**Response**

Dog groups did not show significantly different levels of response to the self and other group's barking (Contingency chi-square test, $\chi 2 = 0.066708$, df = 1, p = 0.796, **Figure 1**). The response and no-response levels were comparable in the self (Goodness of fit, $\chi 2 = 0.1$, df = 1, p = 0.998) and other (Goodness of fit, $\chi 2 = 0.3$, df = 1, p = 0.584) conditions, which suggests that the chance of responding to a bark was random. The latency to respond was comparable in the self and other conditions (Wilcoxon Paired-Sample Test, V = 464, p = 0.305, **Figure 2a**). No significant difference was found in case of response duration in self and other conditions (Wilcoxon Paired-Sample Test, V = 481, p = 0.207, **Figure 2b**). We found no significant difference in the proportion of responsive individuals between self and other conditions (Goodness of Fit, $\chi 2 = 0.00036364$, df = 1, p = 0.984, **Supplementary Figure 3**).

Males and females responded equally in both the self and other conditions (Contingency chi-square test, $\chi 2 = 0.0001$, df = 1, p = 0.997). When the responses were pooled across conditions to check for any gender bias, the overall responses of males (49.5%) and females (50.5%) were comparable (Goodness of fit; $\chi 2 = 0.02$, df = 1, p = 0.887, **Supplementary Figure 4**) suggesting that there was no gender bias in response to the barking sound.

The nature of response in the self and other conditions was comparable (Contingency chi-square test, $\chi 2 = 2.1469$, df = 3, p = 0.542, **Figure 3**). Change of state was the most common of the four kinds of responses (Goodness of fit; $\chi 2 = 29.448$, df = 3, p < 0.005).

**Consistency in trials**

a) **Group level:** The consistency of groups in all the three trials was compared for self and other conditions. There was no significant difference in the consistency for different conditions (Contingency chi-square test, $\chi2 = 3.3929$, df = 1, p = 0.065, **Figure 4**). The observed consistency was not significantly different from the expected consistency, as calculated using the probabilities of response occurring in two of three and three of three trials (Goodness of fit; $\chi2 = 0.10667$, df = 1, p = 0.744).

b) **Individual level:** The consistency of individuals in all the three trials was compared for self and other conditions. There was no significant difference in the consistency for different conditions (Contingency chi-square test: $\chi2 = 0.80585$, df = 1, p = 0.369, **Figure 4**). The observed consistency was comparable to the expected consistency (Goodness of fit: $\chi2 = 2.1572$, df = 1, p = 0.142).

c) **First responder**: The consistency of the first responder in all the three trials was compared for self and other conditions. There was no significant difference in the consistency for different conditions (Contingency chi-square test: $\chi2 = 0.53706$, df = 1, p = 0.463, **Figure 4**). The observed consistency was not significantly different from the expected consistency (Goodness of fit; $\chi2 = 0.88889$, df = 1, p = 0.346). There was no significant difference in the response of the first responder on the basis of gender (Male 52.9%, female 47.1%; Goodness of fit; $\chi2 = 0.29412$, df = 1, p = 0.588).

**Response Index**

The RI was calculated both at the individual and group levels for each type of playback conditions. For each group, we checked for any correlation between the response index of the group and the level (frequency per hour per individual) of aggression shown by the group during observations, as calculated from the AOS data, and found them to be uncorrelated (Pearson's correlation test: df = 38, p = 0.913, $R^2 = 0.178$, **Supplementary Figure 5**).

Similarly, the RI was calculated for each individual, and this was tested against the frequency per hour of aggression behaviour, and the two were found to be uncorrelated (Pearson's correlation test, df = 155, p = 0.552, $R^2 = 0.047$, **Supplementary Figure 5**).

These results suggest that the response to the barking by either the group or an individual is not dependent on the level of aggression by them.

**Sound analysis**

**Group-wise -** The six acoustic parameters for each track were subjected to PCA. PC1 and PC2 together explained 84.54% variation in the data **(See Supplementary Table 1 and Table 2 for details).** The PCA plot revealed clustering of the tracks, irrespective of group identity **(Figure 6a)**.

**Area-wise -** PC1 and PC2 together explained 64.59% variation in the data. The PCA plot revealed clustering of the tracks, irrespective of group's area identity **(Figure 6b)**.

GLMM analysis revealed no significance of the group's size and groups' locality on the RI of the groups **(See Supplementary Table 3 for details).**

The acoustic parameters across groups were mostly overlapping. GLMM analysis with acoustics parameters suggests that Peak_f, Duration 90%, and Entropy have significant effects on the RI of the group **(See Supplementary Table 4 for details).** Using GLM analysis, we found neither group ID nor the track combinations to affect the RI of the group **(See Supplementary Table 5 for details)**.

**Behavioural Observations**

There was no significant difference in the levels of vocalization (Kruskal-Wallis test, $\chi 2 = 3.9574$, df = 3, p = 0.266), and urine marking behaviour (Kruskal-Wallis test, $\chi 2 = 1.0489$, df = 3, p = 0.789). However, aggression in the recording phase and first session of playback were significantly higher than in the second and third sessions of the playback, although aggression in the recording phase was comparable to the first session of the playback (Kruskal-Wallis test, $\chi 2 = 9.3204$, df = 3, p = 0.025) **(See Supplementary Table 6 for details).** Similarly, affiliation behaviour in the recording phase and first session of playback were significantly higher than the third session of the playback and were significantly lower than the second session of the playback, although aggression in the recording phase was comparable to the first session of the playback (Kruskal-Wallis test, $\chi 2 = 8.9904$, df = 3, p-value = 0.029) **(See Supplementary Table 7 for details) (Supplementary Figure 6)**.

**Discussion**

Free-ranging dogs live in stable social groups and are known to be territorial. They show interesting cooperation-conflict dynamics within the group, including parental care, alloparenting, parent-offspring conflict, milk theft, and food sharing (Paul, Sau, Nandi, & Bhadra, 2017)Paul, Majumder, & Bhadra, 2014,Paul & Bhadra, 2017). Such interactions involve communication using visual, auditory, tactile, and olfactory modes. Barking is the most common of all dog vocalizations and is used in various contexts, including territory maintenance, to communicate both within and between groups. Free-ranging dogs exchange barks, growls and howls, for both long and short distance exchanges between individuals and groups. It is common to hear them use vocalizations, especially during the night, when one or more individuals of a group respond to calls of others, who might not be present in the vicinity of the group, or even within the visual range. However, in such cases, the presence of the group members in their territory provides a context to this behaviour. The current study was designed to test whether dogs are capable of distinguishing between the barks of their own group and others when the sounds are played to them in a context-independent scenario.

In our study, the free-ranging dogs showed similar responses to recorded barks of their own and other groups. Not only did the response rates not vary, but the latency to respond, duration of response and nature of response were comparable between the responses to the two kinds of barks. This strongly suggests that the dogs were not able to distinguish between their own group's bark from another group's bark. The response, when observed, mostly consisted of alertness, rather than more energy-intensive responses like movement and vocalizations. This again suggests that the dogs were not much perturbed by the playback sound of the bark that was presented to them out of context and did not have any other kinds of cues, either visual or olfactory, associated with them. This result was consistent at both individual and group levels, and there was no gender bias in the responses, suggesting that the territorial response *per se* is not gender-dependent. This directly challenges the common belief that the alpha member of dog groups is always a male, who is the most aggressive and reactive member of the group. The fact that the dogs responded similarly to all the three sound tracks suggests that they were unable to identify their own barking as distinct from the others.

Our spectrogram analysis of the 400 barks revealed that the barks were extremely similar in their auditory traits, largely overlapping with each other in the PCA space. Thus, it is impossible to distinguish between the different barks analytically. This result, when coupled with the responses of the dogs in the playback experiment, is interesting as it suggests an essential role of context for their behavior. It thus seems reasonable that the dogs were not able to distinguish between the different barks when no context or additional cue was provided to them. Perhaps a combination of visual, olfactory and auditory cues is required to produce territorial responses to the barking of other groups, as opposed to their own. Just the barking sound, detached from any context, produces an immediate response, mostly of alert. The fact that they showed comparable aggression and vocalization levels in the three observation sessions highlights that their responses were momentary and did not stress out the groups or leave an impact on them.

In conclusion, our study points towards the importance of context and combined cues for eliciting behavioral responses from the free-ranging dogs. The presence of humans also makes them cautious. Hence, some biases may have been involved due to the presence of the experimenter within the territory, though this would be a constant bias across experiments. We tried to record the barking tracks at the group's natural habitat, which included background noise while recording. These background noises might interfere with the barking, which makes some biases in response. Despite all these factors, we get an opportunity to understand the response of the free-ranging dogs to different types of barking sounds and from this basic study; many puzzles can be solved like the role of vocalizations in inter and intra-group communications and qualitative analysis of different kinds of dog vocalizations. This study opens up several interesting questions. Since dogs do use barks for communication, it would be interesting to test if barking produced in different contexts elicit different responses from groups/individuals in similar playback experiments. In this study, we have recorded barking randomly and often got the barking produced by a group, rather than an individual. It would thus be interesting to test if the free-ranging dogs can distinguish between barks from an individual versus from a group. Free-ranging dogs are one of the most urban-adapted species. It would be interesting to study how urban noise might have altered the behavior of dogs.

**Acknowledgements**

The experiment was performed by PM for her MS thesis project. PM carried out the data analysis. AB designed the experiment, supervised the work and co-wrote the paper with PM. We thank Dr. Laurel B. Symes (Cornell Lab of Ornithology) for her inputs on the bioacoustics analysis. PM would like to thank Mrs. Alpana Manna for helping in video recording the experiments. PM was supported by INSPIRE fellowship from DST India.

**Conflict of interest statement**

All the authors have read and agree with this version of the manuscript. The authors declare no conflict of interest.

**Ethical statement**

No dogs were harmed during this study. The methods reported in this paper were approved by the animal ethics committee of IISER Kolkata (approval number: 1385/ac/10/CPCSEA), and in accordance with approved guidelines of animal rights regulations of the Government of India.

**Figures:**

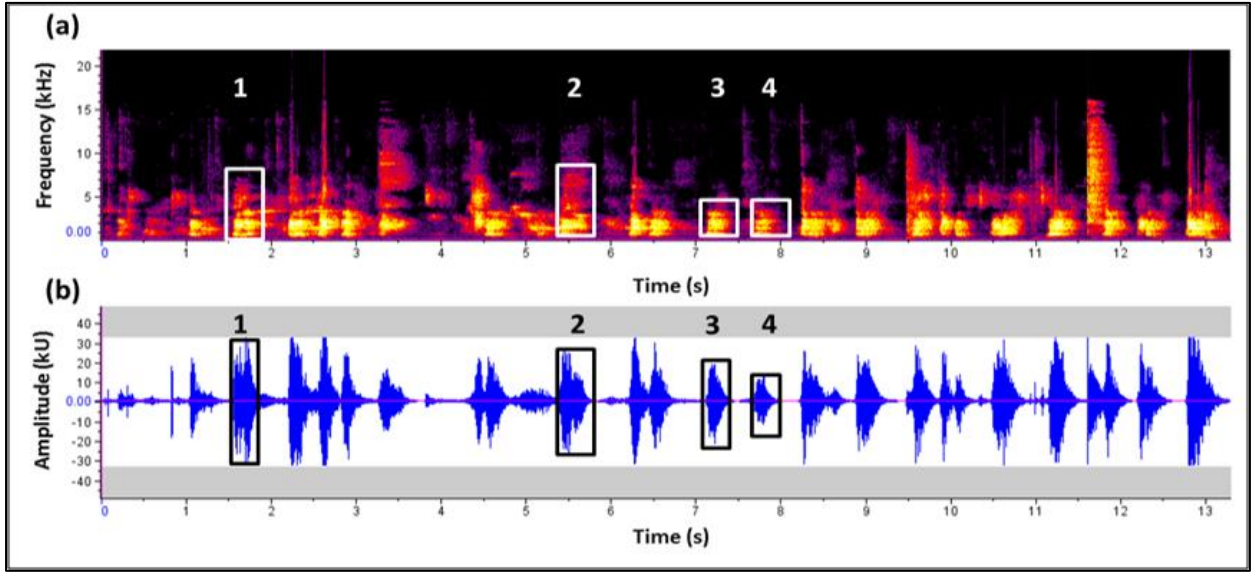

**Figure 1:** Example of a spectrogram and selected syllables from group 5

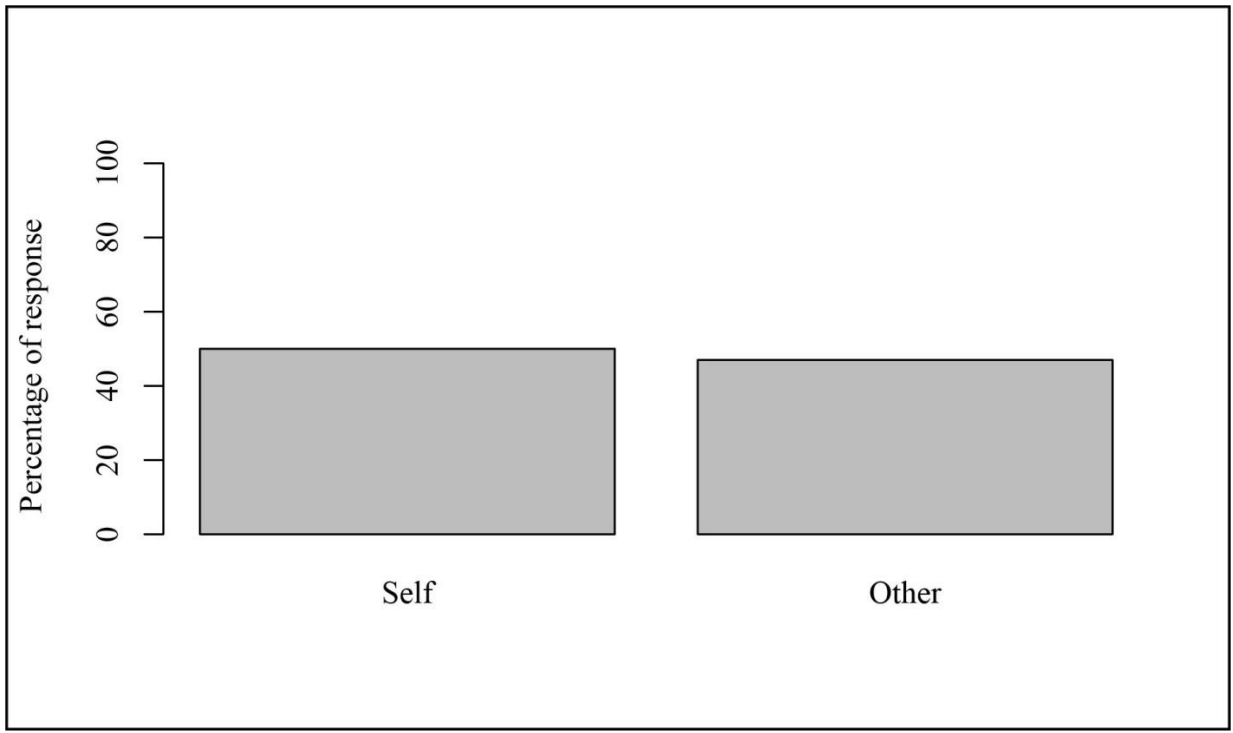

**Figure 2:** A bar graph showing the percentage of dogs which showed a response for the self and other conditions in the playback experiment

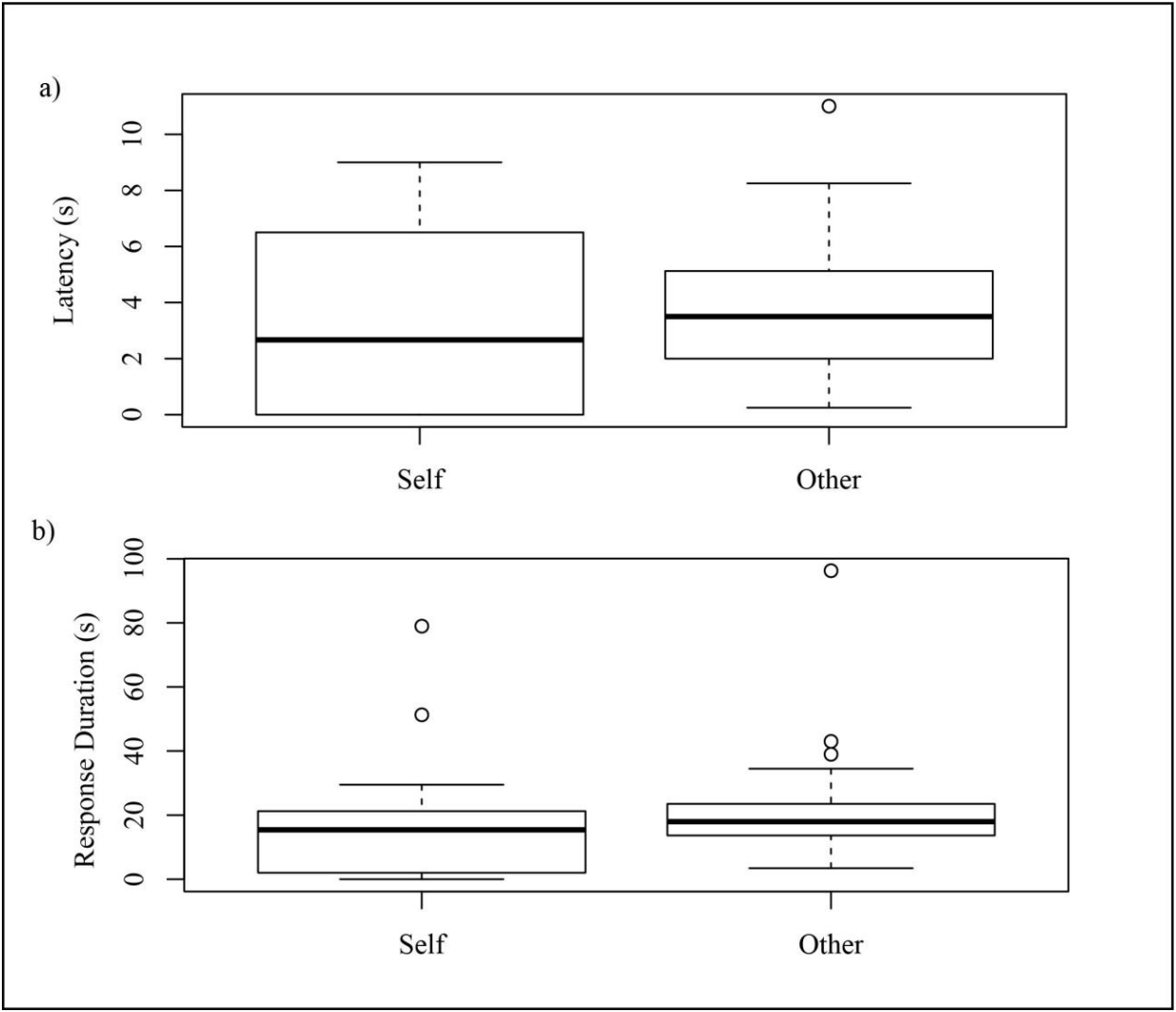

**Figure 3:** Box and whiskers plots showing a) the latency to response (in seconds) and b) the response duration (in seconds) in the two conditions

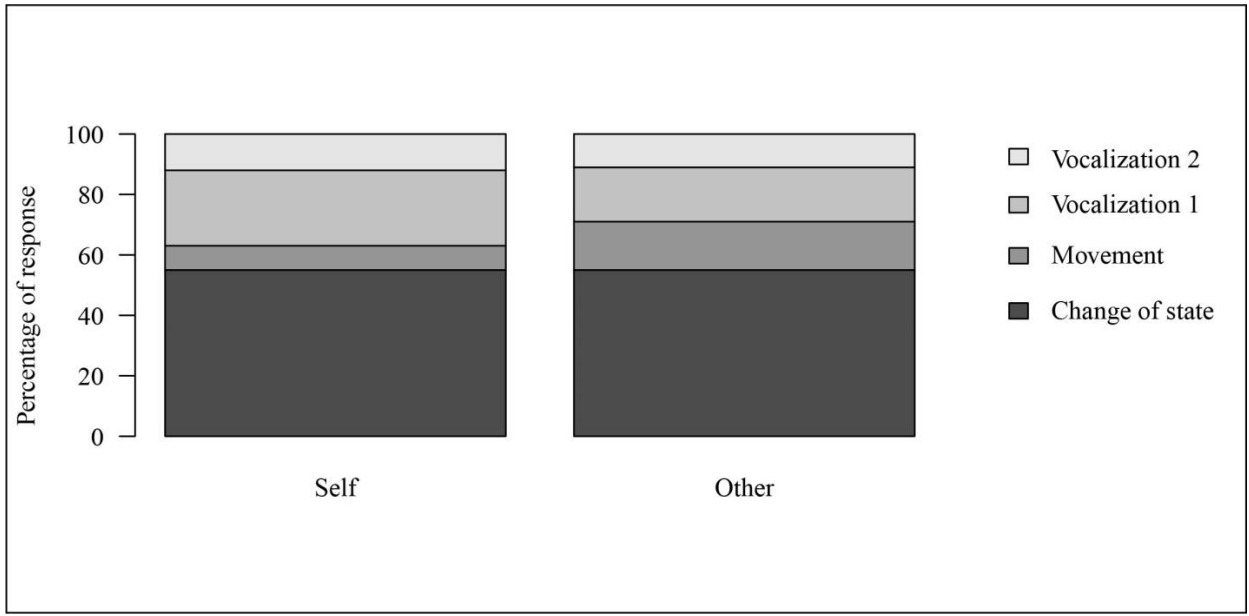

**Figure 4:** A stacked bar graph showing the types of responses in the self and other conditions

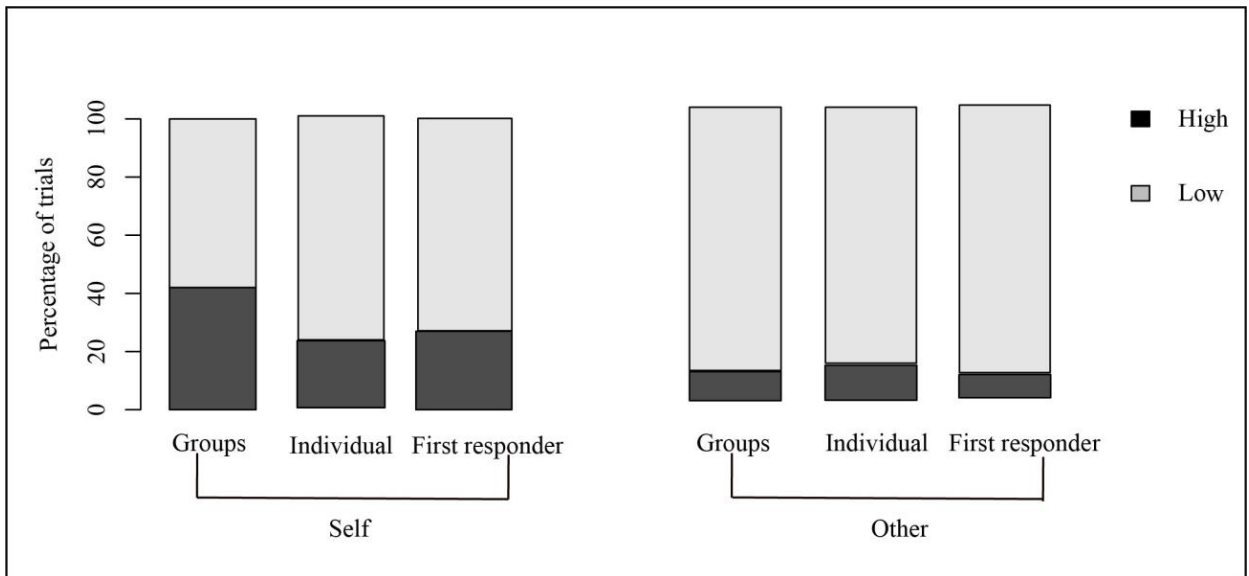

**Figure 5:** Stacked bar graph showing the consistency of groups, individual and the first responder in trials for different conditions

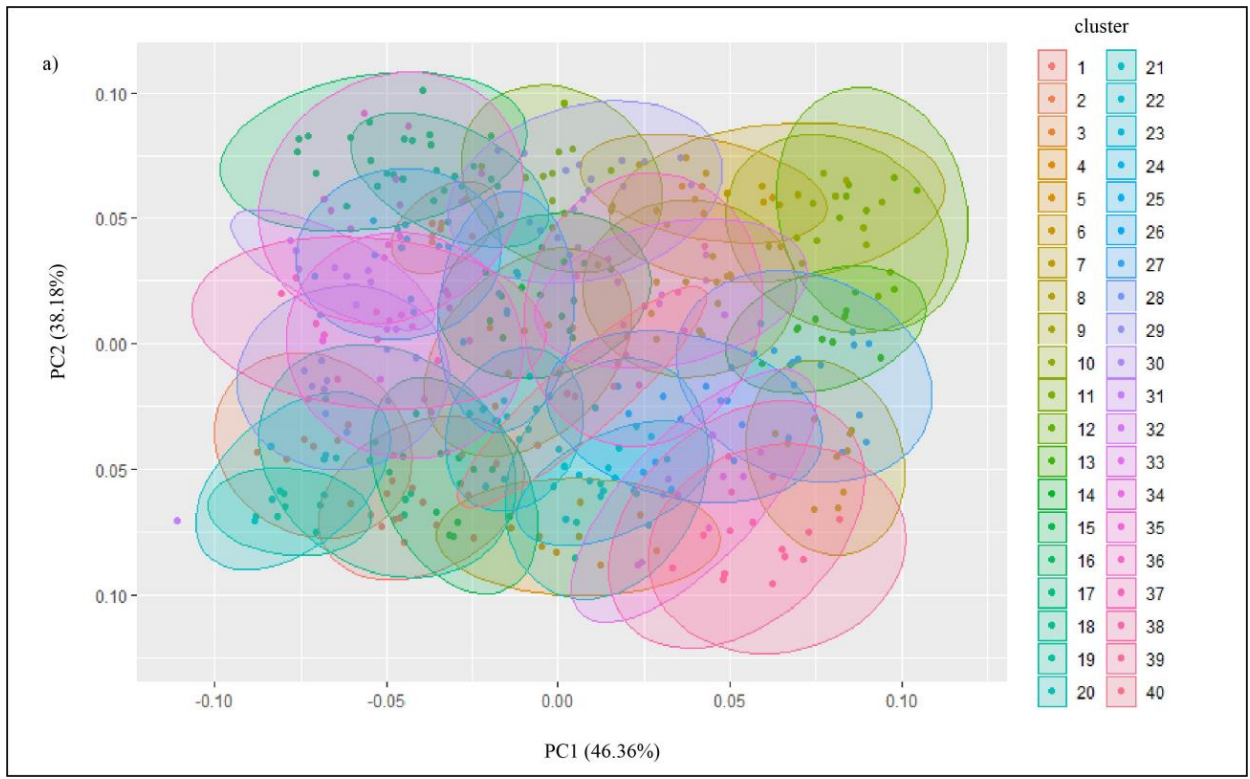

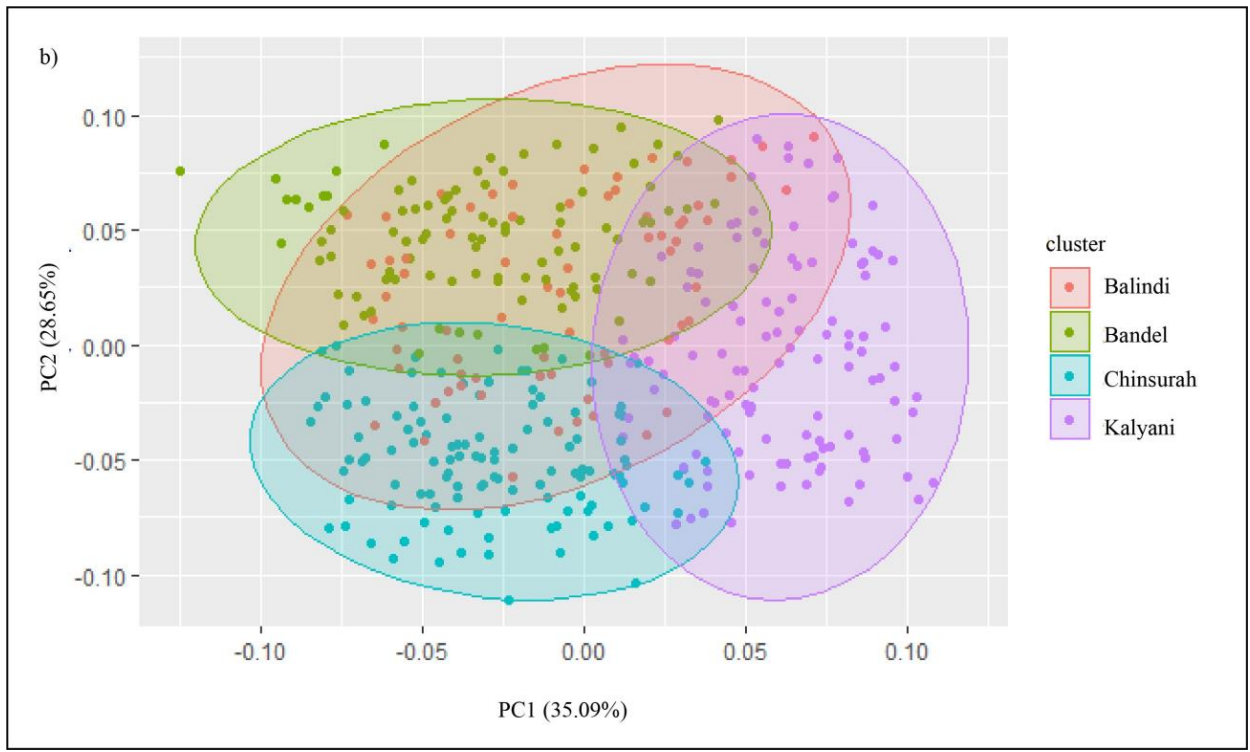

**Figure 6:** PCA plot for all barking tracks plotted using the first two principal components for the syllables from each track (n = 400 syllables). a) different color clusters indicate syllables from

each group, b) different color clusters indicate syllables from each locality

**Supplementary Figures:**

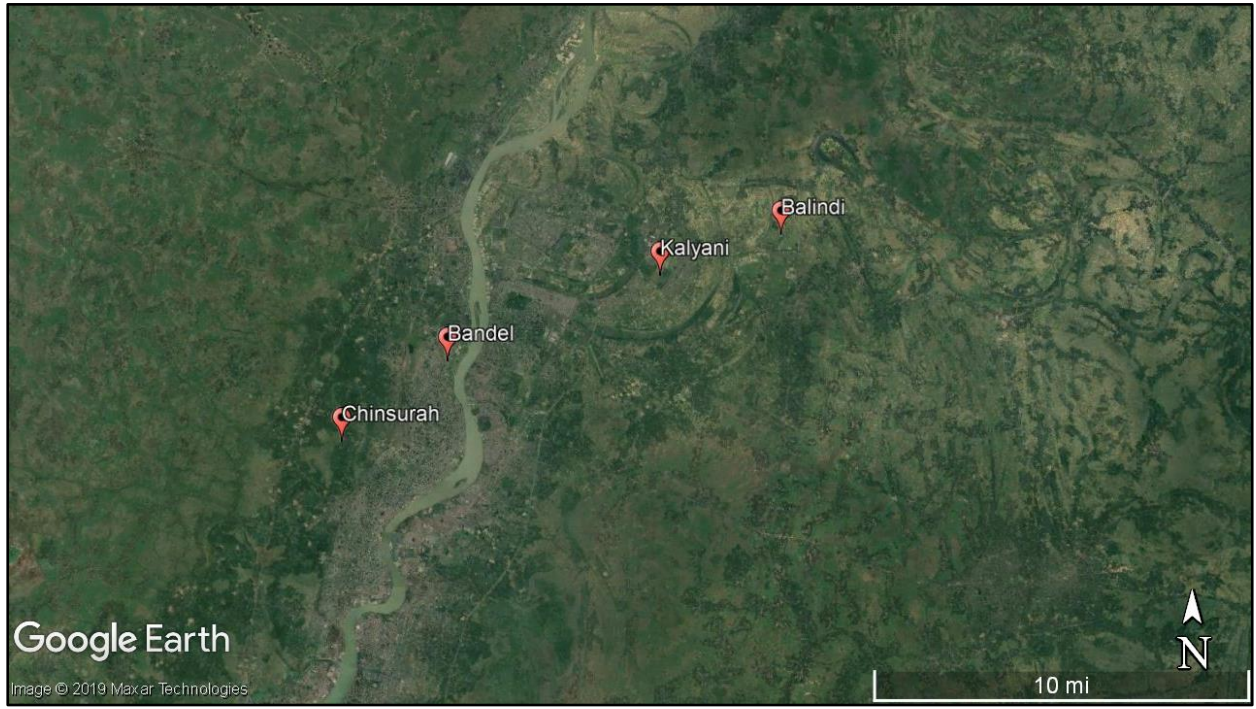

**Figure 1**: Maps showing the four areas, in which the experiment was conducted. The map has been prepared in Google Earth ©

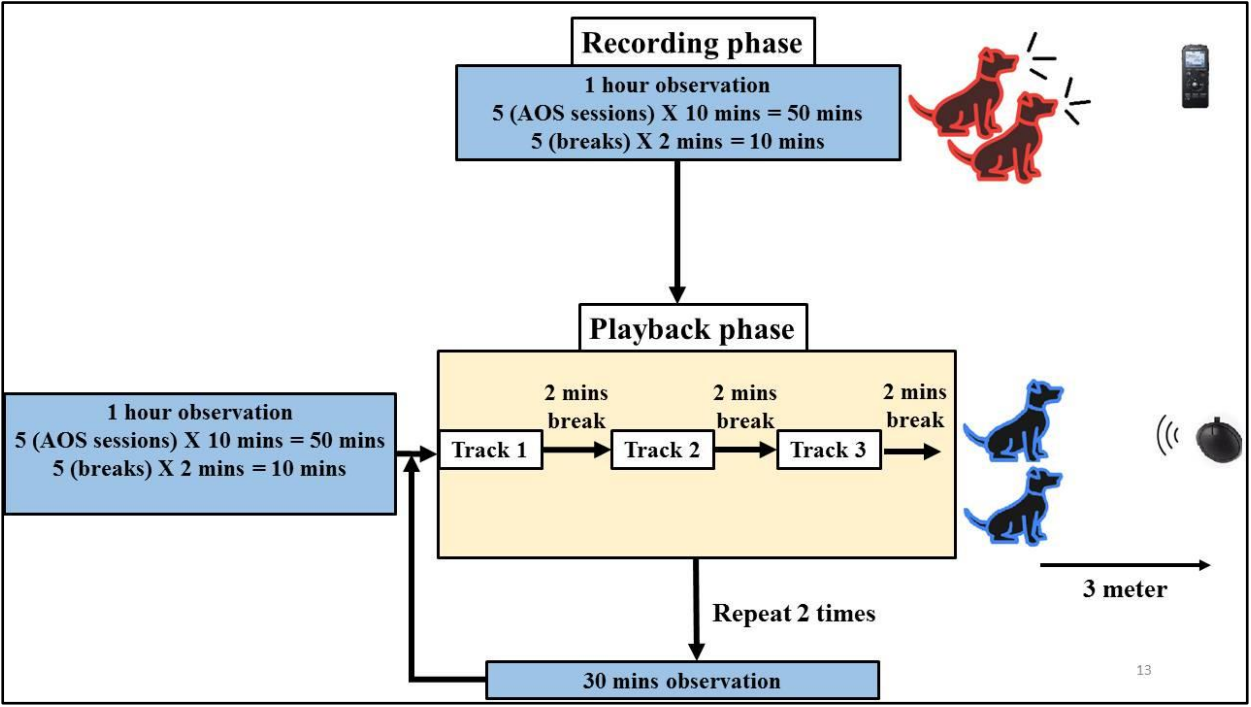

**Figure 2:** A schematic diagram of the protocol

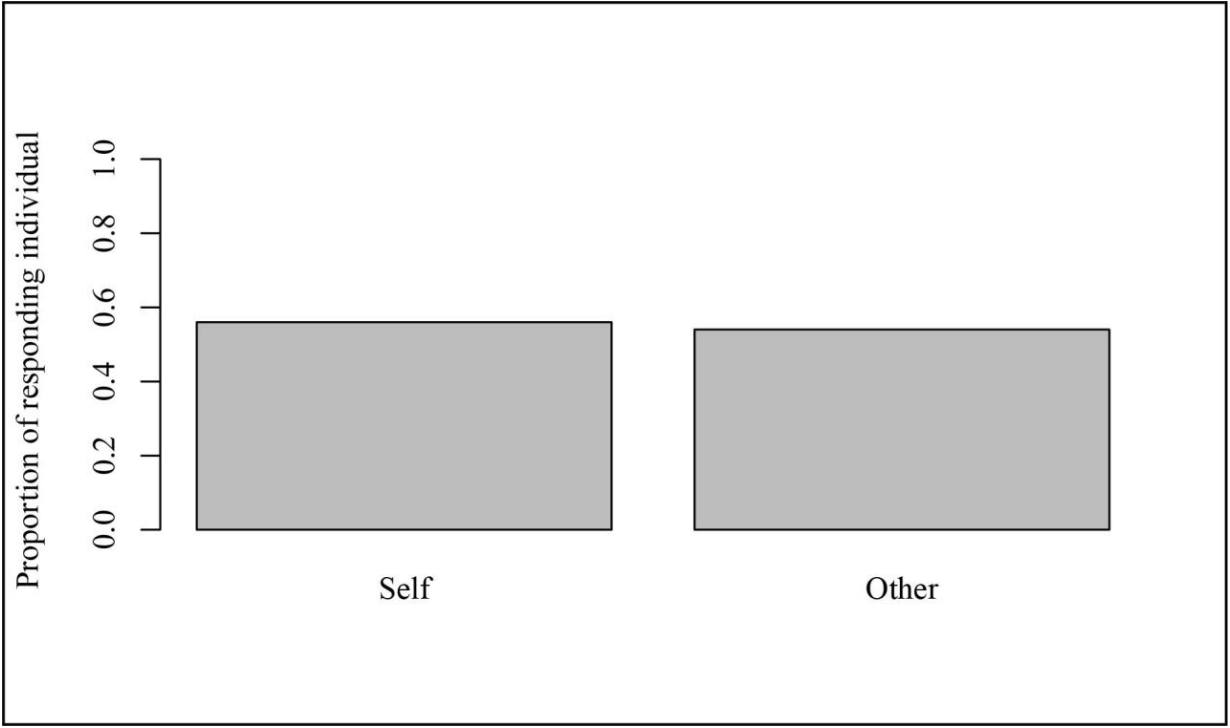

**Figure 3:** A bar graph showing the proportion of responders in a group in the two conditions

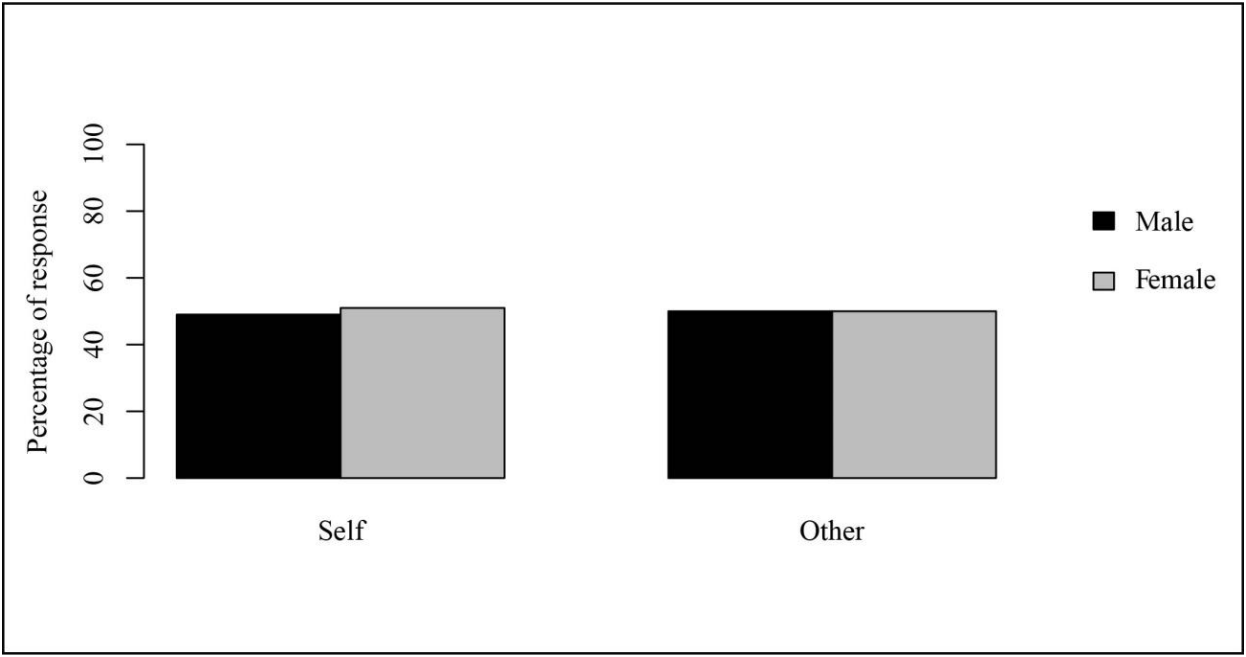

**Figure 4:** A bar graph showing the percentage of response by male (black bars) and female (gray bars) individuals in the self and other conditions

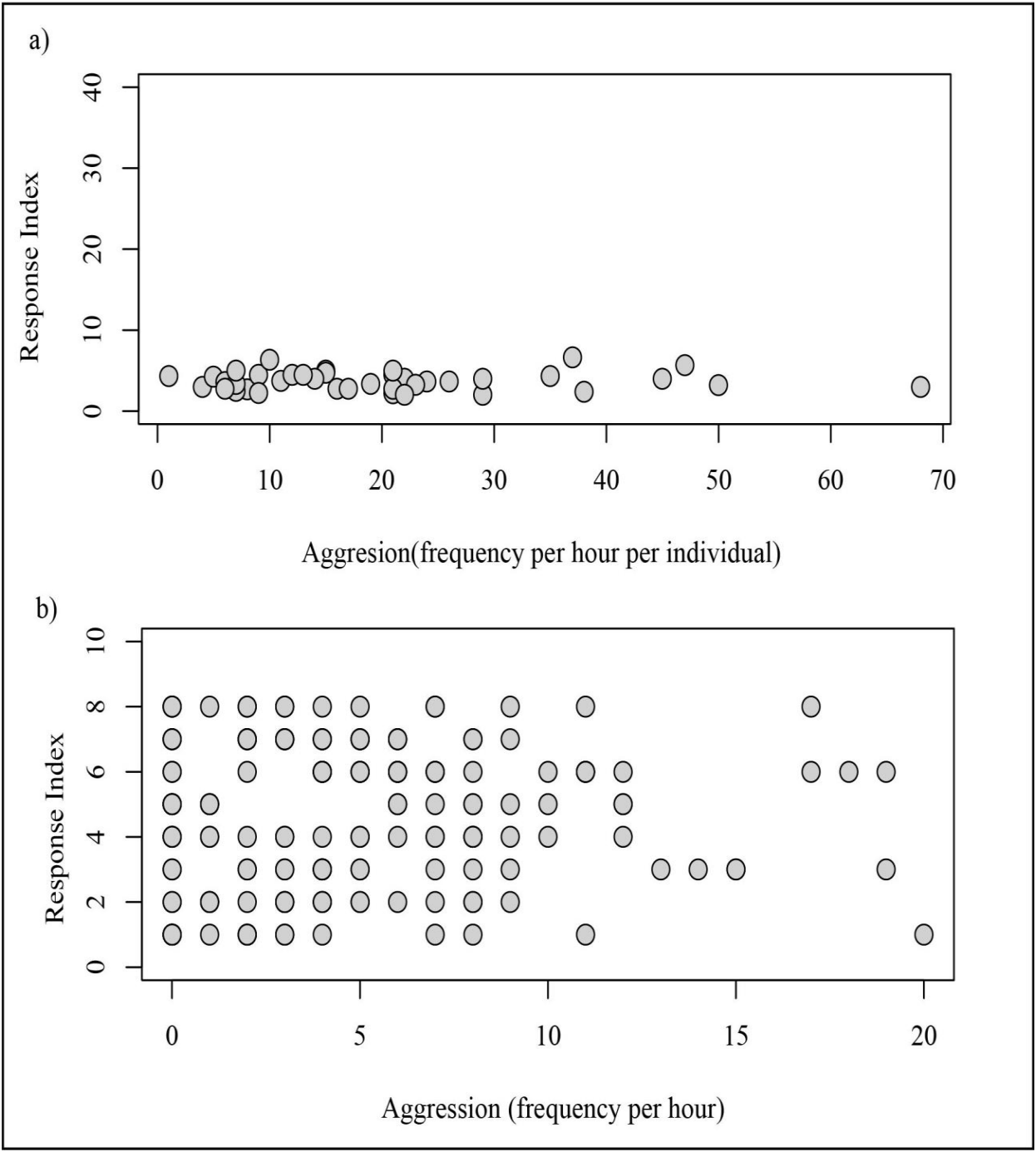

**Figure 5:** Scatter plots showing the correlation between response index and frequency per hour of aggression of a) groups and b) individuals

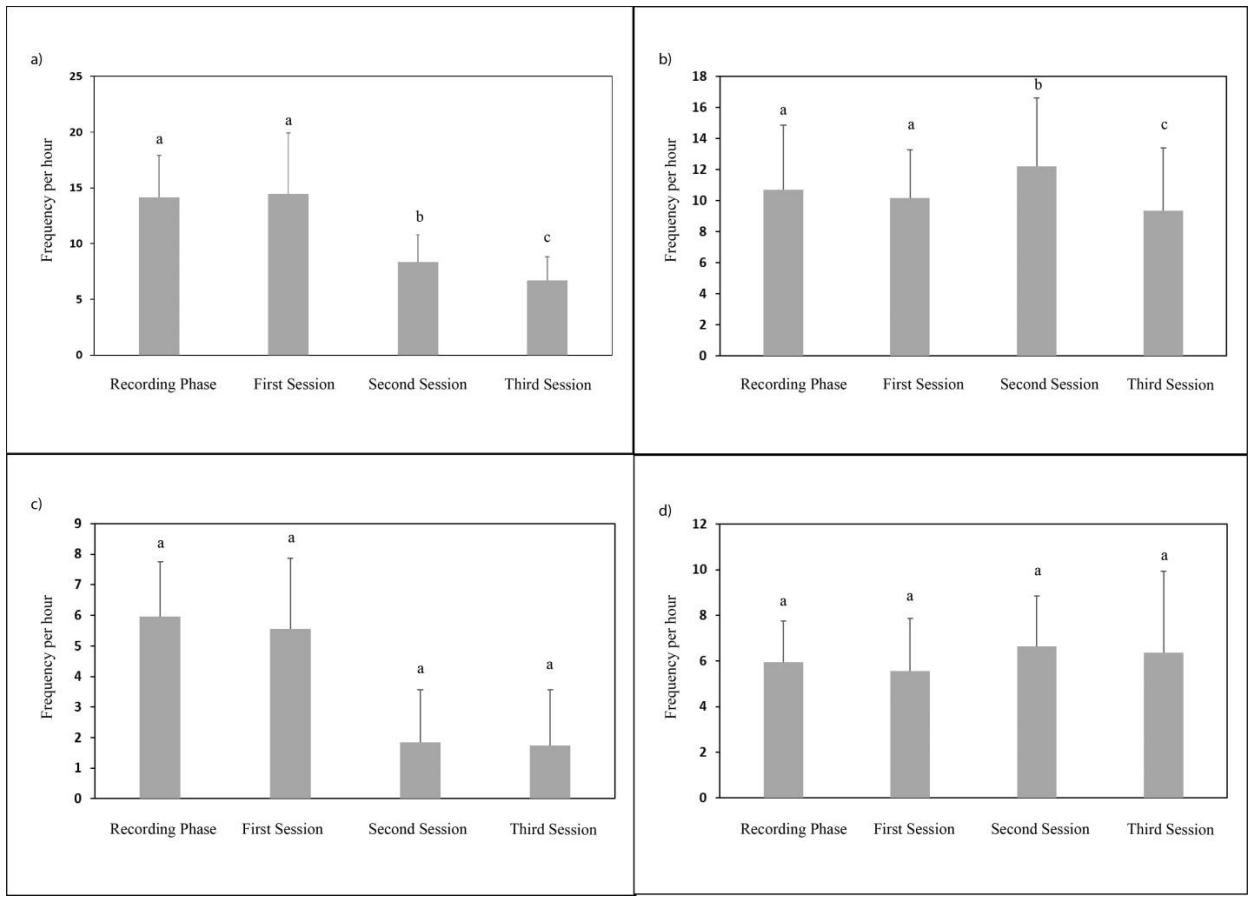

**Figure 6:** Bar graphs showing the mean and standard deviation of frequency per hour of (a) aggression, (b) affiliation, (c) urine marking and (d) vocalization behaviours at the group level in recording phase and playback phase AOS sessions

**Supplementary Tables:**

|  | **PC1** | **PC2** | **PC3** | **PC4** | **PC5** | **PC6** |
|---|---|---|---|---|---|---|
| **Standard deviation** | 419.830 | 380.977 | 228.225 | 81.099 | 0.422 | 0.114 |
| **Proportion of Variance** | 0.464 | 0.382 | 0.137 | 0.017 | 0.000 | 0.000 |
| Cumulative Proportion | 0.464 | 0.846 | 0.983 | 1.000 | 1.000 | 1.000 |

Table 1. Variation explained by the different components of the PCA

| | Comp1 | Comp2 | Comp3 | Comp4 | Comp5 | Comp6 |
|---|---|---|---|---|---|---|
| **High_f** | 0.426 | -0.042 | 0.310 | -0.798 | -0.281 | -0.062 |
| **Low_f** | -0.372 | -0.540 | 0.439 | -0.077 | 0.088 | 0.602 |
| **Peak_f** | 0.070 | 0.550 | 0.512 | -0.027 | 0.651 | 0.065 |
| **Duration 90%** | -0.391 | 0.231 | -0.582 | -0.558 | 0.254 | 0.281 |
| **Bandwidth 90%** | 0.674 | 0.032 | -0.272 | 0.182 | 0.055 | 0.659 |
| **Aggregate Entropy** | -0.259 | 0.591 | 0.189 | 0.106 | -0.649 | 0.339 |

Table 2. Correlations of acoustic parameters with the principal components of the principal component analysis

| Fixed Effects | | | | |
|---|---|---|---|---|
| | Estimate | Std. Error | z value | Pr(>|z|) |
| (Intercept) | 1.035 | 1.178 | 0.879 | 0.379 |
| Group_size | -0.186 | 0.297 | -0.625 | 0.532 |
| LocalityBandel | -2.229 | 1.484 | -1.502 | 0.133 |
| LocalityChinsurah | -1.329 | 1.668 | -0.797 | 0.425 |
| LocalityKalyani | -1.011 | 1.550 | -0.652 | 0.514 |
| Group_size:LocalityBandel | 0.634 | 0.371 | 1.707 | 0.088 |
| Group_size:LocalityChinsurah | 0.400 | 0.420 | 0.953 | 0.341 |
| Group_size:LocalityKalyani | 0.345 | 0.390 | 0.886 | 0.375 |
| Random effects | | | | |
| Groups | Name | Variance | Std.Dev. | |
| Trial:Group_ID | (Intercept) | 0.256 | 0.506 | |
| Group_ID | (Intercept) | 0.296 | 0.544 | |

*Signif. codes:  0 „***" 0.001 „**" 0.01 „*" 0.05 „." 0.1 „ " 1

**Table 3: Table summarising the results from the GLMM analysis with group size and locality.**

| Fixed Effects | | | | |
|---|---|---|---|---|
| | Estimate | Std. Error | z value | Pr(>\|z\|) |
| (Intercept) | 5.349e+02 | 1.817e+02 | 2.944 | 0.003 ** |
| Peak_f | -3.961e-01 | 1.283e-01 | -3.08 | 0.002 ** |
| Duration | -1.821e+03 | 6.113e+02 | -2.979 | 0.003 ** |
| Entropy | -1.384e+02 | 4.783e+0 | -2.894 | 0.004 ** |
| Peak_f:Duration | 1.345e+00 | 4.312e-01 | 3.119 | 0.002 ** |
| Peak_f:Entropy | 1.025e-01 | 3.369e-02 | 3.041 | 0.002 ** |
| Duration:Entropy | 4.738e+02 | 1.611e+02 | 2.941 | 0.003 ** |
| Peak_f:Duration:Entropy | -3.493e-01 | 1.134e-01 | -3.079 | 0.002 ** |
| Random effects | | | | |
| Groups | Name | Variance | Std.Dev. | |
| Trial:Group_ID | (Intercept) | 0.255 | 0.505 | |
| Group_ID | (Intercept) | 0.375 | 0.612 | |

*Signif. codes: 0 „***" 0.001 „**" 0.01 „*" 0.05 „." 0.1 „ " 1

**Table 4: Table summarising the results from the GLMM analysis with acoustic parameters**

| | Estimate | Std. Error | z value | Pr(>\|z\|) |
|---|---|---|---|---|
| (Intercept) | 0.780 | 0.072 | 10.822 | <2e-16 *** |
| Group_ID | 0.002 | 0.003 | 0.529 | 0.597 |

**Table 5: Table summarising the results from the GLM analysis with group's ID**

| | Recording Phase | First session | Second session |
|---|---|---|---|
| First session | U = 840, p = 0.703 | | |
| Second session | U = 1434, p < 0.008 | U = 1303, p < 0.008 | |
| Third session | U = 1562.5, p < 0.008 | U = 1422.5, p < 0.008) | U = 1098, p < 0.008 |

**Table 6: Table summarising Mann-Whitney test results with Bonferroni correction for aggressive behaviour**

|  | Recording Phase | First session | Second session |
|---|---|---|---|
| First session | U = 750.5, p = 0.636 |  |  |
| Second session | U = 1310.5, p < 0.008 | U = 1356, p < 0.008 |  |
| Third session | U = 1439, p < 0.008 | U = 1459, p < 0.008 | U =1096, p = 0.004 |

**Table 7: Table summarising Mann-Whitney test results with Bonferroni correction for affiliative behaviour**